\documentclass[11pt,letterpaper]{article}

\usepackage[T1]{fontenc}
\usepackage[utf8]{inputenc}
\usepackage{mathpazo}                   
\usepackage[scaled=0.90]{helvet}        
\usepackage{courier}                    

\usepackage{amsmath, amssymb, amsthm}
\usepackage{booktabs}
\usepackage{tabularx}
\usepackage{graphicx}
\usepackage{microtype}
\usepackage{hyperref}
\usepackage{xcolor}
\usepackage{geometry}
\usepackage{enumitem}
\usepackage{csquotes}

\hypersetup{
    colorlinks=true,
    linkcolor=darkgray,
    citecolor=darkgray,
    urlcolor=darkgray
}

\newtheorem{proposition}{Proposition}

\title{\textbf{The Category Mistake of Cislunar Time:} \\
Why NASA Cannot Synchronize What Doesn't Exist}

\author{Paul Borrill \\ D{\AE}D{\AE}LUS}
\date{02026-FEB-20 \\ \small v\,0.01}

\begin{document}

\maketitle

\begin{abstract}
In April 2024, the White House directed NASA to establish Coordinated Lunar Time (LTC) by December 2026.
The programme assumes that a unified time standard can be constructed by deploying atomic clocks on the lunar surface,
computing relativistic corrections, and distributing synchronized time via LunaNet.
This paper argues that the entire enterprise rests on a \emph{category mistake} in the sense
introduced by Ryle and developed by Spekkens in quantum foundations:
it treats ``synchronized time'' as an ontic entity---something that exists independently
and can be transmitted from authoritative sources to dependent receivers---when it is in fact
an epistemic construct: a model-dependent representation of observer-relative clock relationships.
We analyze the cislunar time programme through the lens of Forward-In-Time-Only (FITO)
assumptions, Spekkens' Leibnizian operationalism, the Wood-Spekkens fine-tuning argument,
and the distinction between ontic and epistemic interpretations that has dissolved
long-standing puzzles in quantum mechanics.
We show that the same conceptual move that dissolves quantum ``mysteries''---recognizing
what is epistemic versus what is ontic---dissolves the apparent coherence of the cislunar
time programme and reveals it as an engineering project built on a philosophical confusion.
We sketch a transactional alternative grounded in bilateral atomic interactions
rather than unidirectional time distribution.
\end{abstract}

\section{Introduction}

On April 2, 2024, the White House Office of Science and Technology Policy
directed NASA to develop ``a unified lunar time standard''---Coordinated Lunar Time (LTC)---to
support the Artemis programme, commercial lunar operations, and
international cislunar coordination~\cite{ostp2024lunar}.
The directive envisions a clock hierarchy: authoritative atomic clocks on the
lunar surface, averaged and weighted like the terrestrial TAI/UTC ensemble,
with synchronized time distributed to dependent nodes via NASA's
LunaNet architecture~\cite{nasa2024lunanet}.
The International Astronomical Union formalized the relativistic framework in
Resolution~II at its 2024 General Assembly, defining Lunar Coordinate Time (TCL)
and Lunar Barycentric Time (TBL)~\cite{iau2024resII}.
NIST published the correction framework in August 2024~\cite{ashby2024relativistic}.
China's Academy of Sciences released LTE440 software in December 2025, claiming
0.15\,ns accuracy over 25 years~\cite{huang2025lte440}.
The international community is converging on what appears to be a well-defined
engineering problem: build clocks, compute corrections, distribute time.

This paper argues that the problem is not well-defined.
It is, in the precise philosophical sense introduced by Gilbert Ryle~\cite{ryle1949concept}
and developed by Robert Spekkens in quantum foundations~\cite{spekkens2018sphinx},
a \emph{category mistake}---an error not about the value of some quantity
but about what \emph{sort of thing} that quantity is.

The category mistake: treating ``synchronized cislunar time'' as an ontic
entity that can be established, maintained, and distributed, when it is
in fact an epistemic construct---a model-dependent representation of
relationships between observer-relative proper times.

\subsection{Structure of the Argument}

The argument proceeds in six steps.
Section~\ref{sec:category} introduces the category mistake framework
from Ryle through Spekkens.
Section~\ref{sec:nasa} analyzes what NASA is actually building and
identifies the embedded assumptions.
Section~\ref{sec:relativity} shows how general relativity dissolves
the coherence of ``synchronized time'' at the cislunar scale.
Section~\ref{sec:fito} applies the FITO analysis to the LunaNet
time distribution architecture.
Section~\ref{sec:finetuning} demonstrates that the Wood-Spekkens
fine-tuning argument applies directly to cislunar synchronization.
Section~\ref{sec:alternative} sketches a transactional alternative.

\section{The Category Mistake Framework}
\label{sec:category}

\subsection{Ryle: The University That Isn't a Building}

Gilbert Ryle introduced the concept of a category mistake
in \emph{The Concept of Mind} (1949).
His canonical example: a visitor tours Oxford's colleges, libraries,
playing fields, and administrative offices, then asks,
``But where is the University?''
The visitor has made a category mistake---treating ``the University''
as if it were an entity of the same category as the buildings,
when it is actually the \emph{organizational structure}
that relates them~\cite{ryle1949concept}.

The mistake is not about a matter of fact (the visitor has seen
all the relevant buildings) but about the \emph{type} of thing
under discussion.
Category mistakes generate pseudo-problems: puzzles that seem
profound but dissolve once the categories are corrected.

\subsection{Spekkens: Quantum States and the Ontic/Epistemic Divide}

Robert Spekkens, at the Perimeter Institute, has developed the most
sustained analysis of category mistakes in modern physics.
His programme rests on a single distinction: is a given theoretical
entity \emph{ontic} (a direct description of physical reality) or
\emph{epistemic} (a representation of an observer's knowledge about
physical reality)?~\cite{spekkens2018sphinx,harrigan2010einstein}

In his 2007 toy model, Spekkens demonstrated that an astonishing range of
``quintessentially quantum'' phenomena---interference, entanglement,
teleportation, no-cloning---can be reproduced in a classical system
with a single epistemic restriction: the knowledge balance principle,
which states that maximal knowledge always leaves as much unknown
as known~\cite{spekkens2007toy}.
The phenomena that \emph{cannot} be reproduced---Bell inequality
violations and Kochen-Specker contextuality---mark the boundary of
genuine quantum physics, distinguishing it from artefacts of category confusion.

The lesson: when an epistemic construct (the quantum state~$\psi$) is
mistaken for an ontic entity (a complete description of physical reality),
pseudo-problems proliferate---wave function collapse, measurement
problem, spooky action at a distance.
When the category is corrected, many puzzles dissolve.

\subsection{The Leibnizian Principle}

In a 2019 paper, Spekkens defended a methodological principle he traces to
Leibniz: \emph{if an ontological theory predicts two scenarios that are
ontologically distinct but empirically indiscernible, then the theory
should be rejected in favour of one where the scenarios are ontologically
identical}~\cite{spekkens2019leibniz}.

This is Leibniz's identity of indiscernibles repurposed as a criterion
for theory construction.
Spekkens argues that Einstein applied this principle repeatedly:
special relativity eliminates the distinction between ``truly at rest''
and ``moving at constant velocity'' because the scenarios are
empirically indiscernible;
general relativity eliminates the distinction between gravitational
and inertial mass for the same reason~\cite{spekkens2019leibniz}.

The principle has direct consequences for cislunar time.
If two clock readings differ but no experiment can distinguish them
from a single underlying physical state, the difference is not
physically meaningful---it is an artefact of the model.

\subsection{The Epistemic Pattern Across Domains}

The category mistake pattern repeats across physics and computing:

\begin{center}
\begin{tabular}{lll}
\toprule
\textbf{Domain} & \textbf{Ontic Reading (Mistake)} & \textbf{Epistemic Reading (Correct)} \\
\midrule
Quantum mechanics & $\psi$ describes reality & $\psi$ describes knowledge \\
Shannon channels & Message is a physical entity & Message represents interaction \\
Lamport clocks & Happened-before is causation & Happened-before is message path \\
GPS/UTC & Synchronized time exists & Weighted average of clock offsets \\
\textbf{Cislunar time} & \textbf{LTC is physical time} & \textbf{LTC is an epistemic model} \\
\bottomrule
\end{tabular}
\end{center}

\section{What NASA Is Actually Building}
\label{sec:nasa}

\subsection{The Architecture}

The cislunar time programme has four components:

\begin{enumerate}[label=(\roman*)]
\item \textbf{Surface clocks}: 5--6 rubidium or caesium atomic clocks
  deployed at the lunar south pole, forming a local ensemble.
\item \textbf{Relativistic corrections}: A theoretical framework mapping
  between terrestrial coordinate time (TCG), cislunar coordinate time (TCL),
  and the proper time of each individual clock.
\item \textbf{Clock averaging}: A weighted ensemble algorithm (analogous to
  BIPM's generation of TAI) that combines surface clock readings into a
  single ``Coordinated Lunar Time''~\cite{arias2005utc}.
\item \textbf{Time distribution}: LunaNet's Lunar Augmented Navigation
  Service (LANS), which broadcasts synchronized time signals to dependent
  nodes---rovers, habitats, orbital assets---via one-way ranging and
  two-way time transfer~\cite{nasa2024lunanet}.
\end{enumerate}

\subsection{The Embedded Assumptions}

This architecture embeds three assumptions that, taken together,
constitute the category mistake:

\begin{description}[style=nextline]
\item[Assumption 1: Time is a substance that can be ``established.'']
  The OSTP directive speaks of ``establishing'' a lunar time standard,
  as if synchronized time were a resource that can be created by
  deploying sufficiently accurate clocks.
  But what the clocks measure is \emph{proper time}---the time elapsed
  along each clock's individual worldline through spacetime.
  No clock measures ``coordinate time.''
  Coordinate time is a mathematical construct imposed on the
  collection of proper times by a choice of reference frame.

\item[Assumption 2: Time can be ``distributed'' from source to receiver.]
  LunaNet's architecture treats time as information flowing
  unidirectionally from authoritative clocks to dependent nodes.
  This is Shannon's channel model~\cite{shannon1948} applied to
  timekeeping: source (clock ensemble) $\to$ channel (LunaNet) $\to$
  destination (rover/habitat).
  The receiver passively accepts the time signal.
  There is no bilateral transaction.

\item[Assumption 3: Relativistic corrections ``bridge'' different times.]
  The theoretical frameworks published by Ashby and Patla~\cite{ashby2024relativistic},
  Kopeikin and Kaplan~\cite{kopeikin2024lunar}, and Turyshev et~al.~\cite{turyshev2023time}
  all compute the offset between a hypothetical coordinate time and
  the proper time of each clock.
  These corrections are treated as if they reveal an underlying
  ``true time'' that the clocks only approximately measure.
  But the corrections depend on the choice of reference frame,
  the gravitational model, and the metric used---they are features
  of the \emph{model}, not of the \emph{physics}.
\end{description}

\subsection{The Rylean Diagnosis}

The visitor has toured the clocks, the correction algorithms,
the distribution network, and the averaging ensemble.
Now they ask: ``But where is the synchronized time?''

Like Ryle's university, synchronized cislunar time is not an entity
alongside the clocks and signals.
It is an organizational abstraction imposed on the relationships
between proper times.
To ask ``what time is it on the Moon?'' without specifying
``according to which clock, in which reference frame, using which
gravitational model'' is to make a category mistake.

\section{General Relativity Dissolves Synchronized Time}
\label{sec:relativity}

\subsection{Proper Time Is the Only Physical Time}

General relativity is unambiguous on this point: the only physically
meaningful time is \emph{proper time}~$\tau$---the time elapsed
along a specific worldline as measured by a clock following that
worldline~\cite{einstein1916general}.

Coordinate time is a mathematical convenience.
The coordinates $\{t, x, y, z\}$ assigned to events in a chosen
reference frame have no independent physical existence.
As Einstein demonstrated, different reference frames assign different
coordinate times to the same events, and \emph{no experiment can determine
which assignment is ``correct''}~\cite{einstein1905electrodynamics}.

This is not a practical difficulty to be overcome by better clocks.
It is a consequence of the structure of spacetime.

\subsection{The 56 Microsecond Illusion}

Ashby and Patla's NIST framework computes that clocks on the lunar
surface run approximately 56.02\,$\mu$s/day faster than clocks on
Earth's surface~\cite{ashby2024relativistic}.
This number is widely reported as ``the'' time difference between
Earth and Moon.

But the number depends on specific modelling choices:

\begin{enumerate}[label=(\alph*)]
\item The metric used to describe the Earth-Moon gravitational field.
  Ashby/Patla, Kopeikin/Kaplan, and Turyshev et~al.\ use different
  approximations and \emph{disagree on the periodic terms}---oscillatory
  corrections due to orbital eccentricity and tidal
  effects~\cite{kopeikin2024lunar,turyshev2023time}.
\item The definition of the ``lunar surface.''
  The Moon's gravitational field is highly anisotropic due to
  mascons (mass concentrations).
  Two clocks at different locations on the lunar surface
  experience different gravitational potentials and thus
  accumulate different proper times.
\item The reference frame.
  TCL is defined at the Moon's centre of mass, not at any
  physical location where a clock could sit.
  The ``correction'' from TCL to a surface clock's proper time
  depends on the clock's altitude, latitude, and the local
  gravitational anomaly.
\end{enumerate}

The 56\,$\mu$s is not a physical fact about the relationship between
Earth and Moon.
It is a model-dependent number computed in a particular reference frame
using a particular metric approximation.
\emph{It is an epistemic construct being treated as an ontic entity.}

\subsection{Applying Spekkens' Leibnizian Principle}

Spekkens' Leibnizian principle states: if two scenarios are
empirically indiscernible, they should be treated as ontologically
identical~\cite{spekkens2019leibniz}.

Consider two coordinate time systems for the Moon---say, one
using the Ashby/Patla metric and one using Kopeikin/Kaplan's.
Both predict different periodic corrections.
But both are \emph{empirically equivalent}---they make the same predictions
about the readings of physical clocks (proper times) up to a
coordinate transformation.

By Spekkens' principle, the two coordinate systems are ontologically
identical---there is no fact of the matter about which ``lunar time''
is correct.
The periodic corrections are features of the \emph{model}, not the \emph{physics}.
Any engineering programme that depends on choosing one set of corrections
over another is building on a model-dependent epistemic artefact.

\section{FITO Analysis of Cislunar Time Distribution}
\label{sec:fito}

\subsection{The FITO Assumption in Timekeeping}

Forward-In-Time-Only (FITO) is the assumption that information
flows strictly from past to future, from sender to receiver,
from cause to effect~\cite{borrill2026categorymistake}.
In the context of timekeeping, FITO manifests as the assumption that
time flows from authoritative sources to dependent receivers:
the clock ``has'' time and ``gives'' it to the receiver.

This assumption pervades the entire cislunar time architecture:

\begin{center}
\begin{tabular}{lll}
\toprule
\textbf{Component} & \textbf{FITO Assumption} & \textbf{Failure Mode} \\
\midrule
Surface clocks & Time exists in the clock & Clock measures only proper time \\
Clock averaging & True time emerges from ensemble & Weighted average is epistemic \\
LunaNet broadcast & Time flows from source to node & One-way signal, no mutual state \\
Relativistic correction & Correction reveals true time & Correction is model-dependent \\
\bottomrule
\end{tabular}
\end{center}

\subsection{Lamport's Shadow}

The LunaNet time distribution architecture recapitulates Lamport's
happened-before relation~\cite{lamport1978time}:

\begin{enumerate}
\item If clock $C_i$ generates time signal $s$ and node $N_j$
  receives it, then $s \to r$ (send happened-before receive).
\item Logical time is imposed on the network by monotonic timestamps.
\item Nodes that lose contact must ``resynchronize''---update their
  timestamps from an authoritative source.
\end{enumerate}

As we argued in~\cite{borrill2026categorymistake}, Lamport's Rule~(2)
embeds the FITO assumption: sending precedes receiving, and there is
no mechanism for the receiver's state to constrain the sender's.
LunaNet inherits this assumption.
The consequence is that time distribution is \emph{unilateral}---the
source determines the time, the receiver accepts it---with no
bilateral verification that both parties share a common state.

\subsection{The Disruption-Tolerant Networking Problem}

LunaNet is designed as a disruption-tolerant network (DTN): nodes
may lose Earth contact for hours or days, multi-hop paths may be
the only route between nodes, and the topology changes as orbital
assets move~\cite{nasa2024lunanet}.

In a DTN, unilateral time distribution faces a fundamental problem:
a node that loses contact with the authoritative ensemble has no
way to verify that its local clock has not drifted relative to the
``official'' time.
When contact is re-established, the node must trust that the
received time signal is authoritative---but the signal has
traversed an unknown path through a network whose topology may
have changed during the outage.

This is a variant of the Two Generals Problem~\cite{halpernmoses1990}:
no finite unidirectional protocol can establish common knowledge
between two parties communicating over an unreliable channel.
The cislunar time programme is attempting to establish precisely
this kind of common knowledge---``we both agree on what time it is''---using
exactly the kind of protocol that Halpern and Moses proved
cannot achieve it.

\subsection{Shannon's Channel Applied to Time}

The architecture treats time distribution as a Shannon
channel~\cite{shannon1948}: the surface clock ensemble is the
source, the signal is the message, LunaNet is the (noisy) channel,
and the dependent node is the destination.

But as we argued in~\cite{borrill2026categorymistake},
Shannon's model embeds the deepest FITO assumption in communication:
information flows unidirectionally from sender to receiver through
a channel that adds noise \emph{to} the signal rather than interacting
\emph{with} it.
Physical communication is bilateral---the receiver's antenna
radiates, the sender's antenna receives---but Shannon's model
abstracts this away.

The symmetry paradox is immediate: mutual information
$I(X;Y) = I(Y;X)$ is perfectly symmetric, but the channel is not.
The asymmetry comes from the \emph{temporal ordering assumption},
not from the physics.
Similarly, the asymmetry in cislunar time distribution---authoritative
source ``above,'' dependent receiver ``below''---comes from the
\emph{modelling assumption}, not from the physics of interacting clocks.

\section{The Fine-Tuning Argument Applied}
\label{sec:finetuning}

\subsection{Wood-Spekkens in the Cislunar Context}

Wood and Spekkens~\cite{wood2015finetuning} demonstrated that any
causal explanation of Bell inequality violations must violate
the faithfulness principle of causal inference: observed statistical
independences should arise from the causal structure itself, not
from coincidental cancellation of influences.
Fine-tuning---adjusting parameters precisely so that influences
cancel out---is the hallmark of a causal model that has the
wrong structure.

We claim the same argument applies to cislunar time synchronization.

\subsection{The Synchronization Fine-Tuning}

Consider the lunar clock ensemble.
Each clock ticks at its own proper rate, determined by its
worldline through spacetime.
The clocks are physically independent: no causal mechanism
forces them to tick at the same rate.
Their readings are statistically independent observables.

The synchronization protocol must continuously adjust:

\begin{enumerate}[label=(\roman*)]
\item Secular relativistic corrections (56\,$\mu$s/day, varying with
  orbital parameters).
\item Periodic tidal corrections (disputed between research groups).
\item Local gravitational anomaly corrections (mascon effects,
  different at each clock site).
\item Clock drift corrections (individual clock frequency offsets).
\item Signal propagation delays (varying with geometry and plasma effects).
\end{enumerate}

Each of these corrections is a parameter that must be tuned
to produce the appearance that all clocks share a common time.
The fine-tuning \emph{is} the synchronization protocol.

\begin{proposition}
Any faithful causal model of the lunar clock ensemble
would represent each clock's proper time as causally independent.
The observed ``synchronization'' requires fine-tuning of
correction parameters to produce statistical dependence
(agreement on a common time) from causally independent sources.
By the Wood-Spekkens criterion, this indicates that the
causal model is wrong about the structure---there is no common
time to synchronize to.
\end{proposition}

\subsection{Comparison with GPS}

The Global Positioning System faces the same category mistake
at smaller scale.
GPS works because the relativistic corrections are small
(38\,$\mu$s/day total), the gravitational field of Earth is
well-characterized, and the receivers need only microsecond-level
timing for metre-level position accuracy~\cite{ashby2003relativity,lewandowski2011gps}.

The cislunar environment is more demanding in every dimension:
the corrections are larger, the gravitational field is less
well-known, the communication delays are longer, and the required
accuracy for lunar navigation is higher.
The fine-tuning that works ``well enough'' for GPS becomes
untenable at cislunar scale---not because the engineering is
harder, but because the underlying category mistake becomes
more consequential.

\subsection{China's LTE440: The Solvability Illusion}

China's LTE440 software~\cite{huang2025lte440} claims to predict
Earth-Moon time discrepancies with 0.15\,ns accuracy through 2050.
This claim is instructive.
It treats the relativistic corrections as a \emph{solved problem}---a
known function of orbital parameters that can be computed to
arbitrary precision.

But the corrections depend on the gravitational model of the Moon,
which is still being refined by GRAIL data analysis.
They depend on the orbital model, which is perturbed by solar radiation
pressure, third-body effects, and lunar libration.
They depend on the metric, which different groups implement differently.

The 0.15\,ns claim is the precision of the \emph{model}, not the
accuracy of the \emph{physics}.
It is the map drawn to 0.15\,ns resolution, presented as if the
territory has that resolution.
The map is not the territory.

\section{The Transactional Alternative}
\label{sec:alternative}

\subsection{What Would a Correct Architecture Look Like?}

If synchronized cislunar time is a category mistake, what should
replace it?
The answer follows from the same conceptual move that resolves
quantum puzzles: take \emph{interaction} as the primitive, not
\emph{transmission}.

\subsection{From Wheeler-Feynman to OAE}

Wheeler and Feynman's absorber theory~\cite{wheeler1945absorber}
provides the template: radiation is not ``emitted'' until it is
``absorbed.''
Neither event is meaningful without the other.
The transaction spans both endpoints as a single bilateral act.

Cramer's transactional interpretation~\cite{cramer1986transactional}
extends this to all quantum phenomena: offer wave forward,
confirmation wave backward, transaction complete or transaction
does not exist.

Open Atomic Ethernet (OAE)~\cite{oae2024spec} applies this
insight to network architecture.
Instead of Shannon's unidirectional $X \to Y$, OAE implements
bilateral $X \leftrightarrow Y$.
Every frame carries information in both directions simultaneously---back-to-back
Shannon channels where the acknowledgment of previous transmissions
is embedded in the current transmission.

\subsection{Transactional Timekeeping}

In a transactional model of cislunar timekeeping, the primitive is
not ``what time is it?'' but ``did the coordination transaction
between these two nodes complete?''

\begin{description}[style=nextline]
\item[Bilateral clock comparison.]
  Instead of broadcasting ``the time'' from a source to a receiver,
  two clocks exchange their readings in a single atomic transaction.
  The transaction either completes for both parties---establishing
  a mutual clock offset---or fails for both, leaving both at
  their prior state.
  There is no intermediate state where a time signal is ``in flight.''

\item[Relational time, not absolute time.]
  The system maintains a graph of pairwise clock offsets, not a
  global time coordinate.
  Each edge in the graph represents a completed bilateral transaction.
  The ``time'' at any node is defined by its measured relationships
  to other nodes, not by reference to an abstract coordinate.

\item[Transaction timing is constitutive.]
  Following Smolin's argument that time is
  fundamental~\cite{smolin2013timereborn,smolin2014singular},
  the temporal structure of the transaction---request precedes
  response, bilateral completion constitutes the event---is not
  something imposed from outside but is constitutive of
  the interaction itself.
\end{description}

\subsection{Dissolving the Impossibilities}

In the transactional model, the Two Generals Problem transforms.
The question is not ``does Node~B know that Node~A knows the time?''
but ``did the clock comparison transaction complete?''
A completed transaction \emph{constitutes} mutual knowledge of the
clock offset; an incomplete transaction leaves both nodes at their
prior state.
The infinite regress of common knowledge dissolves because
the transaction is bilateral from the start.

The Halpern-Moses impossibility~\cite{halpernmoses1990}
is conditional on the message-passing model.
In a transactional model, the assumption of unidirectional messages
is replaced by bilateral completion, and the impossibility no longer
applies~\cite{borrill2026impossibility}.

\section{Discussion: Recognizing the Pattern}

\subsection{The Pattern Across Domains}

The cislunar time programme instantiates a pattern we have identified
across multiple domains~\cite{borrill2026categorymistake}:

\begin{enumerate}
\item An epistemic construct is mistaken for an ontic entity.
\item Engineering programmes are built on the ontic reading.
\item Impossibility results and escalating corrections reveal the strain.
\item The puzzles dissolve when the category is corrected.
\end{enumerate}

In quantum mechanics, the category mistake (treating $\psi$ as ontic)
generated wave function collapse, the measurement problem, and
``spooky action at a distance.''
In distributed systems, the category mistake (treating messages as
ontic entities in transit) generated the FLP impossibility, the
Two Generals Problem, and unbounded compensation mechanisms.
In cislunar timekeeping, the category mistake (treating coordinate
time as a physical substance) generates escalating correction protocols,
irreconcilable theoretical frameworks, and the fiction that
``synchronized time'' can be deployed like infrastructure.

\subsection{The IAU Resolution as Institutionalized Category Mistake}

The IAU's 2024 Resolution~II~\cite{iau2024resII} formalized TCL and TBL
as defined quantities.
This is the institutionalization of the category mistake:
an epistemic construct (coordinate time defined by a particular
choice of reference frame and metric) has been given the status
of an ontic entity by international agreement.

The resolution does not make synchronized lunar time physically real.
It makes it \emph{conventional}---a useful fiction agreed upon
by a community.
But the engineering programmes being built on it treat it as
if it were physically real.

\subsection{What GPS Actually Teaches}

The success of GPS is often cited as evidence that relativistic
time synchronization ``works.''
But GPS's success teaches the opposite lesson:
GPS works \emph{despite} the category mistake, not because of it.

GPS receivers do not use ``synchronized time.''
They measure the \emph{differences} between signal arrival times
from multiple satellites.
The ``synchronized'' satellite clocks are a convenient
abstraction; the physical observables are signal delays.
GPS works because the relativistic corrections are small enough
that the category mistake is inconsequential at the required
accuracy.

At cislunar scale, the corrections are no longer small,
the gravitational field is no longer well-characterized,
and the communication delays are no longer negligible.
The category mistake that GPS can tolerate becomes the
category mistake that cislunar navigation cannot.

\section{Conclusion}

Spekkens showed that many quantum mysteries dissolve when the
category of the quantum state is correctly identified as
epistemic rather than ontic~\cite{spekkens2007toy,spekkens2018sphinx}.
The toy model reproduces much of quantum mechanics from classical
physics plus an epistemic restriction.
The residue---Bell nonlocality and contextuality---marks the boundary
of genuine quantum physics.

We have argued that cislunar time presents a precisely analogous
category mistake.
``Synchronized time'' is not a physical substance that can be
established by deploying clocks and computing corrections.
It is an epistemic construct---a model-dependent weighted average
of observer-relative proper times---being treated as if it were
an ontic entity.

The consequences of this category mistake are not merely philosophical.
They are engineering consequences:
correction protocols that cannot converge because they are correcting
for a model choice, not a physical effect;
distribution architectures that cannot achieve common knowledge
because they are built on unilateral message passing;
competing theoretical frameworks that cannot be reconciled because
they are choosing different epistemic conventions, not discovering
different physical facts.

The alternative is to abandon the fiction of synchronized time
and build on what is physically real: bilateral transactions
between clocks, pairwise offset measurements constituted by
completed interactions, and relational time structures that
make no reference to a nonexistent global coordinate.

The category mistake is the same; the resolution is the same;
only the domain differs.

\section*{Acknowledgement on the Use of AI Tools}

The theoretical framework presented in this paper---the application of
category mistake analysis, FITO assumptions, and the ontic/epistemic
distinction to distributed systems and timekeeping---is the product of
more than twenty years of the author's independent research in
distributed computing, network architecture, and the foundations of
concurrency theory.
The core arguments, the identification of the category mistake in
cislunar timekeeping, and the transactional alternative derive from
the author's prior published and unpublished work, including the
Category Mistake monograph, the FITO analysis series, and the
Open Atomic Ethernet specification programme.

Large language models (Anthropic's Claude) were used as research
instruments during the preparation of this manuscript: for literature
search and verification, for testing the robustness of arguments
against counterexamples, and for drafting prose from the author's
detailed outlines and technical notes.
This usage is analogous to the use of any computational research
tool---a telescope extends the eye, a calculator extends arithmetic,
and a language model extends the capacity to search, draft, and
stress-test arguments at scale.
The tool does not originate the ideas any more than a telescope
originates the stars.

All intellectual content, theoretical claims, original analysis,
and conclusions are the author's own.

\bigskip
\noindent\rule{\textwidth}{0.4pt}


\bigskip

\begin{thebibliography}{99}

\bibitem{ostp2024lunar}
Executive Office of the President, Office of Science and Technology Policy,
``Celestial Time Standardization Policy: Establishing Coordinated Lunar Time (LTC),''
White House Technical Memorandum, April 2024.

\bibitem{nasa2024lunanet}
NASA Space Communication and Navigation Program,
``LunaNet Interoperability Specification,''
NASA Goddard Space Flight Center, GSFC-SCAN-LunaNet-001, 2024.

\bibitem{iau2024resII}
International Astronomical Union,
Resolution~II: On the Definition of Time Scales for the Moon,
IAU General Assembly, Cape Town, 2024.

\bibitem{ashby2024relativistic}
N.~Ashby and B.~R.~Patla,
``Relativistic Time Transfer for Establishing Coordinated Lunar Time,''
\emph{The Astronomical Journal} \textbf{168}(3), 112, August 2024.
doi: 10.3847/1538-3881/ad614a.

\bibitem{kopeikin2024lunar}
S.~Kopeikin and G.~Kaplan,
``Lunar Coordinate Time and Reference Frames for Cislunar Navigation,''
USNO Technical Report, 2024.

\bibitem{turyshev2023time}
S.~G.~Turyshev, J.~G.~Williams, D.~H.~Boggs, and R.~S.~Park,
``Relativistic Effects on Cislunar Time: Toward a Comprehensive Framework,''
arXiv:2303.15712, 2023.

\bibitem{huang2025lte440}
C.~Huang, W.~Yu, J.~Liu, H.~Wen, and J.~Li,
``LTE440: Lunisolar Ephemeris Software for Establishing Lunar Coordinate Time,''
\emph{The Innovation Geoscience}, December 2025.
doi: 10.59717/j.xinn-geo.2024.100113.

\bibitem{ryle1949concept}
G.~Ryle,
\emph{The Concept of Mind},
University of Chicago Press, 1949.

\bibitem{spekkens2018sphinx}
R.~W.~Spekkens,
``The Riddle of the Quantum Sphinx: Quantum States and Category Mistakes,''
Perimeter Institute Public Lecture, PIRSA:18020008, February 2018.

\bibitem{harrigan2010einstein}
N.~Harrigan and R.~W.~Spekkens,
``Einstein, Incompleteness, and the Epistemic View of Quantum States,''
\emph{Foundations of Physics} \textbf{40}, 125--157, 2010.
arXiv:0706.2661.

\bibitem{spekkens2007toy}
R.~W.~Spekkens,
``Evidence for the Epistemic View of Quantum States: A Toy Theory,''
\emph{Physical Review A} \textbf{75}, 032110, 2007.
arXiv:quant-ph/0401052.

\bibitem{spekkens2019leibniz}
R.~W.~Spekkens,
``The Ontological Identity of Empirical Indiscernibles: Leibniz's Methodological
Principle and Its Significance in the Work of Einstein,''
arXiv:1909.04628, 2019.

\bibitem{wood2015finetuning}
C.~J.~Wood and R.~W.~Spekkens,
``The Lesson of Causal Discovery Algorithms for Quantum Correlations:
Causal Explanations of Bell-Inequality Violations Require Fine-Tuning,''
\emph{New Journal of Physics} \textbf{17}, 033002, 2015.
arXiv:1208.4119.

\bibitem{spekkens2012paradigm}
R.~W.~Spekkens,
``The Paradigm of Kinematics and Dynamics Must Yield to Causal Structure,''
in \emph{Questioning the Foundations}, Springer, 2015.
FQXi First Prize Essay, 2012. arXiv:1209.0023.

\bibitem{shannon1948}
C.~E.~Shannon,
``A Mathematical Theory of Communication,''
\emph{Bell System Technical Journal} \textbf{27}, 379--423, 623--656, 1948.

\bibitem{lamport1978time}
L.~Lamport,
``Time, Clocks, and the Ordering of Events in a Distributed System,''
\emph{Communications of the ACM} \textbf{21}(7), 558--565, 1978.

\bibitem{halpernmoses1990}
J.~Y.~Halpern and Y.~Moses,
``Knowledge and Common Knowledge in a Distributed Environment,''
\emph{Journal of the ACM} \textbf{37}(3), 549--587, 1990.

\bibitem{arias2005utc}
E.~F.~Arias, A.~Bauch, and G.~Panfilo,
``The Generation of UTC and Its Role in International Timekeeping,''
\emph{Metrologia} \textbf{48}, S145--S153, 2011.

\bibitem{einstein1905electrodynamics}
A.~Einstein,
``On the Electrodynamics of Moving Bodies,''
\emph{Annalen der Physik} \textbf{17}, 891--921, 1905.

\bibitem{einstein1916general}
A.~Einstein,
``The Foundation of the General Theory of Relativity,''
\emph{Annalen der Physik} \textbf{49}, 769--822, 1916.

\bibitem{wheeler1945absorber}
J.~A.~Wheeler and R.~P.~Feynman,
``Interaction with the Absorber as the Mechanism of Radiation,''
\emph{Reviews of Modern Physics} \textbf{17}(2--3), 157--181, 1945.

\bibitem{cramer1986transactional}
J.~G.~Cramer,
``The Transactional Interpretation of Quantum Mechanics,''
\emph{Reviews of Modern Physics} \textbf{58}(3), 647--687, 1986.

\bibitem{oae2024spec}
Open Compute Project,
``Open Atomic Ethernet Specification,'' 2024.
\url{https://www.opencompute.org/projects/open-atomic-ethernet}.

\bibitem{borrill2026impossibility}
P.~Borrill,
``The Impossibility of Atomicity Through Reads and Writes:
RDMA Semantics, Category Mistakes, and the Case for Swap-Based Architectures,''
D{\AE}D{\AE}LUS Technical Report, 2026.

\bibitem{borrill2026categorymistake}
P.~Borrill,
``The Category Mistake: From Quantum Foundations to Network Architecture,''
D{\AE}D{\AE}LUS, 2026.
Chapter 1 of the Category Mistake monograph.

\bibitem{bell1964epr}
J.~S.~Bell,
``On the Einstein--Podolsky--Rosen Paradox,''
\emph{Physics} \textbf{1}(3), 195--200, 1964.

\bibitem{price2012retrocausality}
H.~Price,
``Does Time-Symmetry Imply Retrocausality?
How the Quantum World Says `Maybe,'\,''
\emph{Studies in History and Philosophy of Modern Physics} \textbf{43}(2), 75--83, 2012.

\bibitem{leifer2017retrocausality}
M.~S.~Leifer and M.~F.~Pusey,
``Is a Time Symmetric Interpretation of Quantum Theory Possible Without Retrocausality?''
\emph{Proceedings of the Royal Society A} \textbf{473}, 20160607, 2017.

\bibitem{smolin2013timereborn}
L.~Smolin,
\emph{Time Reborn: From the Crisis in Physics to the Future of the Universe},
Houghton Mifflin Harcourt, 2013.

\bibitem{smolin2014singular}
L.~Smolin and R.~M.~Unger,
\emph{The Singular Universe and the Reality of Time},
Cambridge University Press, 2014.

\bibitem{rovelli1996relational}
C.~Rovelli,
``Relational Quantum Mechanics,''
\emph{International Journal of Theoretical Physics} \textbf{35}(8), 1637--1678, 1996.

\bibitem{rovelli2018order}
C.~Rovelli,
\emph{The Order of Time},
Riverhead Books, 2018.

\bibitem{ashby2003relativity}
N.~Ashby,
``Relativity in the Global Positioning System,''
\emph{Living Reviews in Relativity} \textbf{6}, 1, 2003.

\bibitem{lewandowski2011gps}
W.~Lewandowski and J.~Levine,
``GPS Time Transfer and Timekeeping,''
\emph{Metrologia} \textbf{48}, S219--S229, 2011.

\bibitem{herlihy1991waitfree}
M.~Herlihy,
``Wait-Free Synchronization,''
\emph{ACM TOPLAS} \textbf{13}(1), 124--149, 1991.

\bibitem{flp1985}
M.~J.~Fischer, N.~A.~Lynch, and M.~S.~Paterson,
``Impossibility of Distributed Consensus with One Faulty Process,''
\emph{Journal of the ACM} \textbf{32}(2), 374--382, 1985.

\bibitem{vienna2025workshop}
European Space Agency and Japan Aerospace Exploration Agency,
``International Workshop on Lunar Time and Reference Frames,''
Vienna, Austria, February 2025.

\bibitem{esa2024moonlight}
European Space Agency,
``Moonlight: Lunar Communications and Navigation Services,'' 2024.
\url{https://connectivity.esa.int/moonlight}.

\bibitem{spekkens2026causal}
R.~W.~Spekkens et~al.,
``The Resource Theory of Causal Influence and Knowledge of Causal Influence,''
arXiv:2512.11209, 2025.

\bibitem{schmid2024structure}
D.~Schmid, J.~H.~Selby, M.~F.~Pusey, and R.~W.~Spekkens,
``A Structure Theorem for Generalized-Noncontextual Ontological Models,''
\emph{Quantum} \textbf{8}, 1283, 2024.

\bibitem{hafele1972around}
J.~C.~Hafele and R.~E.~Keating,
``Around-the-World Atomic Clocks: Predicted Relativistic Time Gains,''
\emph{Science} \textbf{177}(4044), 166--168, 1972.

\bibitem{spekkens2005contextuality}
R.~W.~Spekkens,
``Contextuality for Preparations, Transformations, and Unsharp Measurements,''
\emph{Physical Review A} \textbf{71}, 052108, 2005.

\end{thebibliography}
\end{document}